\documentclass[a4paper,aps,prd,12pt,superscriptaddress, nofootinbib, prd]{revtex4-2}

\usepackage[utf8]{inputenc}
\usepackage[T1]{fontenc}
\usepackage{bm}
\usepackage{tikz}
\usepackage{amsmath}
\usepackage{amssymb}
\usepackage{graphicx}
\usepackage[colorlinks=true, allcolors=blue]{hyperref}
\usepackage{slashed}
\usepackage{dsfont}
\usepackage[english]{babel}
\usepackage[normalem]{ulem}

\newcommand{\ri}{\mathrm{i}}
\newcommand{\re}{\mathrm{e}}
\newcommand{\rd}{\mathrm{d}}
\newcommand{\U}{\mathrm{U}}
\newcommand{\SU}{\mathrm{SU}}
\usepackage{natbib}

\DeclareMathOperator{\tr}{tr}
\DeclareMathOperator{\Tr}{Tr}

\begin{document}

\title{Vorticity-induced effects from Wess-Zumino-Witten terms}

\author{Geraint W. Evans}
\email{geraint@gate.sinica.edu.tw}
\affiliation{Institute of Physics, Academia Sinica, Taipei 11529, Taiwan}

\author{Naoki Yamamoto}
\email{nyama@rk.phys.keio.ac.jp}
\affiliation{Department of Physics, Keio University , Yokohama 223-8522, Japan}

\author{Di-Lun Yang}
\email{dlyang@phys.sinica.edu.tw}
\affiliation{Institute of Physics, Academia Sinica, Taipei 11529, Taiwan}
\affiliation{Physics Division, National Center for Theoretical Sciences, Taipei 106319, Taiwan}

\begin{abstract}
We study vorticity-induced effects arising from the Wess-Zumino-Witten terms for Nambu-Goldstone modes in chiral perturbation theory. We first provide an alternative derivation of the Wess-Zumino-Witten terms in the presence of external vector, axial-vector, and pseudoscalar fields using a derivative expansion of the fermion determinant. We then employ the previously found correspondence in which vorticity is treated as an axial-vector field coupled to Dirac fermions in flat spacetime. Using this, we derive vorticity-induced contributions for Nambu-Goldstone modes in the presence of electromagnetic fields at finite baryon and isospin chemical potentials, including a vorticity-induced current, a magnetic-field-induced angular momentum, and a vorticity-modified photon–pion coupling. We also briefly discuss the phenomenological implications of these vorticity-induced effects.
\end{abstract}

\maketitle

\section{Introduction}
\label{sec:intro}
Quantum anomalies are among the ways that quantum physics deviates from our classical expectations of nature. A well-known example is the chiral anomaly, where the axial-vector current is no longer conserved once we quantize the classical field theory. This has significant physical consequences, such as contributing to the width of the neutral pion decay into two photons \cite{Adler:1969gk,Bell:1969ts}. 
Since this anomaly is tied to the topological nature of the theory and does not depend on the energy scale, it provides nonperturbative constraints on low-energy physics, known as the ’t Hooft anomaly matching condition \cite{tHooft:1979rat}.

The non-conservation of a non-Abelian axial-vector current in the presence of external fields was first derived by Bardeen~\cite{Bardeen:1969md}, where the divergences of the vector and axial-vector fields are written in a form such that the vector current is conserved. Bardeen's form of the anomaly was later shown to obey certain consistency conditions by Wess and Zumino \cite{Wess:1971yu}, which led to it being called the consistent form of the anomaly. This differentiates it from the covariant anomaly, where both vector and axial-vector currents transform covariantly but neither are conserved in general \cite{Bardeen:1984pm}. 
Wess and Zumino also described some of the low-energy manifestations of the chiral anomaly, including the effective action for Nambu-Goldstone (NG) modes including the $K^{+}K^{-}\rightarrow\pi^+\pi^-\pi^0$ interaction \cite{Wess:1971yu}. This was later extended by Witten \cite{Witten:1983tw}, who showed the anomalous effective action in the absence of external fields can be written as an integral over five spacetime dimensions, now known as the Wess-Zumino-Witten (WZW) term. Shortly after, the full WZW effective action in the presence of external fields was explicitly determined \cite{Kaymakcalan:1983qq,Manohar:1984uq,Kawai:1984mx,Manes:1984gk,Pak:1984bn,Chou_1984,Alvarez-Gaume:1984zlq}.

Since these seminal works, the WZW terms have found numerous applications, including their effects on finite-density QCD matter under strong magnetic fields \cite{Son:2004tq,Son:2007ny,Brauner:2016pko} and/or rotation \cite{Huang:2017pqe,Manes:2019fyw}, in connection with the chiral magnetic effect (CME) \cite{Vilenkin:1980fu,Nielsen:1983rb,Alekseev:1998ds,Fukushima:2008xe} and chiral vortical effect (CVE) \cite{Vilenkin:1979ui,Kharzeev:2007tn,Son:2009tf,Landsteiner:2011cp}. 
The presence of the WZW terms reveals novel QCD phase structures, such as the pion domain wall \cite{Son:2007ny}, chiral soliton lattice (CSL) \cite{Brauner:2016pko}, baryon crystal \cite{Evans:2022hwr,Evans:2023hms}, and domain-wall Skyrmion phases \cite{Eto:2025fkt} in magnetic fields. Analogous phase structures have also been discussed in rotating systems \cite{Huang:2017pqe,Nishimura:2020odq,Eto:2021gyy,Eto:2023tuu}. On the other hand, WZW terms involving rotation or vorticity themselves remain less well explored, apart from Refs.~\cite{Huang:2017pqe,Manes:2019fyw}. For example, the so-called helical magnetic effect (HME) \cite{Yamamoto:2015gzz,Kharzeev:2018jip}, which is a current along the magnetic field in the presence of fluid helicity, is known to be related to the anomaly \cite{Yamamoto:2021gts} and is expected to satisfy anomaly matching. Nevertheless, the corresponding low-energy effective theory has not yet been formulated. In this work, while not addressing the HME itself, we derive previously unexplored vorticity-induced effects within the framework of chiral perturbation theory (ChPT) for NG modes. 

From a phenomenological viewpoint, these vorticity-induced effects may be relevant to relativistic heavy ion collisions, where the strong vorticity is extracted from measurements of global spin polarization of $\Lambda$ hyperons in noncentral collisions \cite{STAR:2017ckg}. 
There have been extensive theoretical studies of QCD matter under rotation, ranging from fundamental properties such as the QCD phase transition \cite{Yamamoto:2013zwa,Jiang:2016wvv,Chernodub:2016kxh,Wang:2018sur,Fujimoto:2021xix,Braguta:2025ddq} to heavy-ion phenomenology, such as thermal dilepton emission \cite{Singh:2018bih,Wei:2021dib,Dong:2021fxn}. 
However, most of these previous studies focus on the deconfined phase, employing lattice QCD with imaginary rotation or phenomenological models.
Meanwhile, the vorticity may remain strong even in the hadronic phase as suggested by the enhanced hyperon polarization in low-energy heavy ion collisions \cite{STAR:2021beb,HADES:2022enx}. It is hence important to establish, from first principles, the low-energy effective action with hadronic degrees of freedom under rotation.  

In this paper, we first present an alternative derivation of the WZW terms in the presence of external vector,
axial-vector, and pseudoscalar fields using a derivative expansion of the fermion determinant following the method of Ref.~\cite{Aitchison:1985pp}. We then employ the correspondence found in Ref.~\cite{Yamamoto:2021gts}, whereby vorticity is treated as an axial-vector field  coupled to Dirac fermions in flat spacetime. 
On this basis, we derive vorticity-induced contributions for NG modes in the presence of electromagnetic fields at finite baryon and isospin chemical potentials, including a vorticity-induced current, a magnetic-field-induced angular momentum, and a vorticity-modified photon–pion coupling.  Although we use the linear sigma model as the underlying microscopic theory, our results should be model-independent.

This work is structured as follows. First, we introduce our linear sigma model in Sec.~\ref{sec:setup} and outline the derivative expansion of fermion determinants.
We then construct the effective action in Sec.~\ref{sec:res} and compare with the already established results. We explore some consequences of these terms in Sec.~\ref{sec:app}. In Sec.~\ref{sec:summary}, we summarize our results and provide further discussion. We work in natural units $\hbar=c=1$ and use the conventions for the metric tensor $\eta_{\mu\nu}={\rm diag}(1,-1,-1,-1)$ and totally anti-symmetric tensor $\epsilon^{\mu\nu\alpha\beta}$ with $\epsilon^{0123}=1$.

\section{Linear Sigma Model and Derivative Expansion of Fermion Determinants}
\label{sec:setup}
Our starting point is the linear sigma model Lagrangian 
\begin{equation}
    \label{lagLSM}
    \begin{split}
        \mathcal{L} = \bar{\psi}\left(\ri\slashed{\partial} -\slashed{V}+\gamma^5\slashed{A}-m \re^{\ri\theta\gamma^5}\right)\psi\,,
    \end{split}
\end{equation}
where $\psi$ is the Dirac spinor field with fermion mass $m$, and $N_f$ flavor and $N_c$ color degrees of freedom.
Here, $V_{\mu}$ and $A_{\mu}$ are respectively the external vector and axial-vector fields. We use ``slash'' notation, $\slashed{A}\equiv\gamma^{\mu}A_{\mu}$ and $\gamma^5 \equiv \ri\gamma^0\gamma^1\gamma^2\gamma^3$. This is identical to the starting point of Ref.~\cite{Yamamoto:2021gts} except we have introduced the pseudoscalar fields $\theta$ following Ref.~\cite{Aitchison:1985pp} to examine the low-energy physics. As it stands, we can accommodate local, non-Abelian vector and axial-vector symmetries provided that the pseudoscalar fields also transform in an appropriate manner. Our initial results will be relevant for such cases. However, we will eventually be interested in Abelian fields only, especially when $A_{\mu}$ can be interpreted as the vorticity. Thus, we can think of the local transformation
\begin{equation}
    \label{locax}
    \begin{split}
        \psi\rightarrow \re^{-\ri\lambda(x)\gamma^5}\psi\,,\qquad A_{\mu}\rightarrow A_{\mu}+\partial_{\mu}\lambda\,, \qquad \phi\rightarrow \phi+2\lambda\,,
    \end{split}
\end{equation}
under which our Lagrangian is invariant when $\theta=\phi$, which could be interpreted as a flavor-singlet field. 
To allow for additional pseudoscalar mesons, we take
\begin{equation}
    \label{decomp}
    \begin{split}
        \theta=\Pi+\phi\,,
    \end{split}
\end{equation}
where $\Pi=\pi_a T_a/f_{\pi}$ is the $\SU(N_f)$ pseudoscalar field which does not transform under the local $\U(1)$ axial symmetry (\ref{locax}). Here, $T_a$ ($a=1,2,\cdots,N_f^2-1$) are the generators of $\SU(N_f)$ and $f_{\pi}$ is the pion decay constant. 

To determine the WZW terms from Eq.~\eqref{lagLSM}, we employ the derivative expansion method of Ref.~\cite{Aitchison:1985pp}. Our partition function is
\begin{equation}
    \label{Z}
    \mathcal{Z} = \int D\psi D\bar{\psi}\, \exp\left\{\ri\int \rd^4 x\left[\bar{\psi}\left(\ri\slashed{\partial} -\slashed{V}+\gamma^5\slashed{A}-m \re^{\ri\theta\gamma^5}\right)\psi\right] \right\}\,,
\end{equation}
and we can integrate out fermions to obtain the effective action
\begin{equation}
    \label{GammaLog}
    \Gamma = -\ri\tr\ln{\left[\slashed{p} -\slashed{V}+\gamma^5\slashed{A}-m \re^{\ri\theta\gamma^5}\right]}\,,
\end{equation}
with momentum operator $p_{\mu}$. Here, ``$\tr$'' denotes the functional trace over the implicit spacetime indices, and the internal spaces, like flavor, color, and Dirac space. From here, we would like to extract the effective Lagrangian $\mathcal{L}_{\text{eff}}$ via
\begin{equation}
    \label{GammaLeff}
    \Gamma = \int \rd^4x\, \mathcal{L}_{\text{eff}}\,.
\end{equation}
This can be achieved to a certain order in the external fields and their derivatives by  expanding around $V_{\mu}=A_{\mu}=\theta=0$, such that
\begin{equation}
    \label{GammaExp}
    \begin{split}
    \Gamma 
    &=
    -\ri\tr\ln{\left(\slashed{p} -m\right)} 
    +\ri\tr\frac{1}{\slashed{p}-m}\tilde{M} 
    +\frac{\ri}{2}\tr\frac{1}{\slashed{p}-m}\tilde{M}\frac{1}{\slashed{p}-m}\tilde{M}
    +\dots\,,
    \end{split}
\end{equation}
where $\tilde{M}=\slashed{V} -\gamma^5\slashed{A}+m(\re^{\ri\theta\gamma^5}-1)$. To place the above effective action in the form of Eq.~\eqref{GammaLeff}, we can use the relation
\begin{equation}
     \label{TrInt}
    \tr\hat{\mathcal{P}} \hat{\mathcal{X}} 
    =\int \rd^4x\,\Tr\langle x | \hat{\mathcal{P}} \hat{\mathcal{X}} | x \rangle 
    =  \int \frac{\rd^4p}{(2\pi)^4} \int \rd^4x  \Tr \mathcal{P}(p) \mathcal{X}(x)\,,
\end{equation}
where $\hat{\mathcal{P}}$ and $\hat{\mathcal{X}}$ are operators with eigenvalues $\mathcal{P}(p)$ and $\mathcal{X}(x)$ respectively. The trace ``$\Tr$'' is no longer over spacetime indices but over the internal spaces only. Before applying this relation, we would first need to separate the position and momentum operators. To do this, we use the commutation relations
\begin{equation}
    \label{Commute1}
    \begin{split}
        [p^{\mu},\varphi] = \ri\partial^{\mu}\varphi \,, 
    \end{split}
\end{equation}
and 
\begin{equation}
    \label{Commute2}
    \begin{split}
    \left[\varphi,\frac{1}{p^2 -m^2}\right] = \frac{1}{(p^2 -m^2)^2}\left[p^2,\varphi\right] + \frac{1}{(p^2 -m^2)^3}\left[p^2,\left[p^2,\varphi\right]\right] + \dots\,,
    \end{split}
\end{equation}
(see, e.g., Ref.~\cite{Novikov:1984ecy} for a proof) with
\begin{equation}
    \begin{split}
    [p^2,\varphi] = \partial_{\mu}\partial^{\mu}\varphi +2\ri p^{\mu}\partial_{\mu}\varphi\,.
    \end{split}
\end{equation}
After this, one can perform the momentum integral in Eq.~\eqref{TrInt} and obtain the result in the form of Eq.~\eqref{GammaLeff}, where the explicit expression of ${\cal L}_{\rm eff}$ will be given in Sec.~\ref{sec:res}.

\section{WZW effective action with external fields}
\label{sec:res}

In this section, we present the results of the derivative expansion up to fourth order in the external fields $V_{\mu}$, $A_{\mu}$, and $\theta$. This leaves many terms to be computed from the expansion \eqref{GammaExp}. The WZW terms are proportional to $\epsilon^{\mu\nu\alpha\beta}$ and one can use this fact to determine which terms will contribute to the effective action. In essence, we only need to consider terms where the total number of $A_{\mu}$ and $\theta$ fields is odd. This follows from the standard Dirac-trace identity, $\Tr_D\gamma^{\mu}\gamma^{\nu}\gamma^{\alpha}\gamma^{\beta}\gamma^{5} = -4\ri\epsilon^{\mu\nu\alpha\beta}$. 
We note that even after applying this selection rule, some terms will still end up vanishing. We do not discuss such terms further. In the following, we also omit the details of each calculation. Interested readers are provided details of one of the calculations in Appendix~\ref{app:deriv} and are referred to Ref.~\cite{Aitchison:1985pp} for further information.
The first nonvanishing terms emerge at third order in the fields. These are the terms with one $\theta$ and two $V_{\mu}$ or $A_{\mu}$. As a shorthand, we label theses terms $\theta VV$ and $\theta AA$ respectively. We will employ similar shorthands below when referring to pieces of the effective action. The $\theta VV$ term comes from considering 
\begin{equation}
    \label{ExpThetaVV}
    \begin{split}
       \ri\tr\frac{1}{\slashed{p}-m}(\ri m\gamma^5\theta)\frac{1}{\slashed{p}-m}\slashed{V}\frac{1}{\slashed{p}-m}\slashed{V}\,,
    \end{split}
\end{equation}
in the expansion \eqref{GammaExp}, where we have expanded $\exp{(\ri\theta\gamma^5)}$ in $\tilde{M}$ for small $\theta$ as in Ref.~\cite{Aitchison:1985pp}. We have included all terms which are equivalent under the cyclicity of the trace in the above, whose addition cancels the factor of $1/3$.
After collecting the momentum operators together and performing the momentum integrals, we obtain the effective action
\begin{equation}
    \label{GammaThetaVV}
    \begin{split}
        \Gamma_{\theta VV} = \frac{1}{8\pi^2}\epsilon^{\mu\nu\alpha\beta}\int \rd^4x\Tr\,\partial_{\mu}\theta\partial_{\nu}V_{\alpha}V_{\beta}\,,
    \end{split}
\end{equation}
up to second order in derivatives. This was previously derived in Ref.~\cite{Aitchison:1985pp} for the case with electromagnetic fields and $\SU(N_f)$ pseudoscalar field. The $\theta AA$ contribution comes from the equivalent term in Eq.~\eqref{ExpThetaVV} with both $V_{\mu}$ replaced by $-\gamma^5A_{\mu}$. We find
\begin{equation}
    \begin{split}
        \Gamma_{\theta AA} 
        = \frac{1}{24\pi^2}\epsilon^{\mu\nu\alpha\beta}\int \rd^4x\Tr\,\partial_{\mu}\theta\partial_{\nu}A_{\alpha}A_{\beta}\,,
    \end{split}
\end{equation}
up to second order in derivatives. 

The simplest nonvanishing fourth-order term is $\theta\theta\theta V$, which can be calculated from 
\begin{equation}
    \label{Exp3ThetaV}
    \begin{split}
        \ri\tr\frac{1}{\slashed{p}-m}(\ri m\gamma^5\theta)\frac{1}{\slashed{p}-m}(\ri m\gamma^5\theta)\frac{1}{\slashed{p}-m}(\ri m\gamma^5\theta)\frac{1}{\slashed{p}-m}\slashed{V}\,,
    \end{split}
\end{equation}
yielding
\begin{equation}
    \label{Gamma3ThetaV}
    \begin{split}
        \Gamma_{\theta\theta\theta V} 
        = -\frac{\ri}{24\pi^2}\epsilon^{\mu\nu\alpha\beta}\int \rd^4x \Tr \,\partial_{\mu}\theta \partial_{\nu}\theta \partial_{\alpha}\theta V_{\beta}\,,
    \end{split}
\end{equation}
up to third order in derivatives. This was also obtained in Ref.~\cite{Aitchison:1985pp} for the case with electromagnetic fields and $\SU(N_f)$ pseudoscalar field. By expanding terms similar to Eq.~\eqref{Exp3ThetaV} but with two of the $\ri m\theta\gamma^5$ factors replaced by $\slashed{V}$, we find the $\theta VVV$ effective action
\begin{equation}
    \label{Exptheta3V}
    \begin{split}
        \Gamma_{\theta VVV} 
        = \frac{\ri}{8\pi^2}\epsilon^{\mu\nu\alpha\beta}\int \rd^4 x \Tr \left(\partial_{\mu}\theta V_{\nu}V_{\alpha} +\theta V_{\nu}\partial_{\mu}V_{\alpha}\right)V_{\beta}\,,
    \end{split}
\end{equation}
up to third order in derivatives. 

Also, the $\theta\theta VA$ contribution arises not only from the fourth-order terms 
\begin{subequations}
    \label{Exp4thThetaThetaVA}
    \begin{align}
        \label{Exp4thThetaThetaVA1}
        \ri\tr\frac{1}{\slashed{p}-m}(\ri m\gamma^5\theta)\frac{1}{\slashed{p}-m}(\ri m\gamma^5\theta)\frac{1}{\slashed{p}-m}\slashed{V}\frac{1}{\slashed{p}-m}(-\gamma^5\slashed{A})\,,
        \\
        \label{Exp4thThetaThetaVA2}
        \ri\tr\frac{1}{\slashed{p}-m}(\ri m\gamma^5\theta)\frac{1}{\slashed{p}-m}(\ri m\gamma^5\theta)\frac{1}{\slashed{p}-m}(-\gamma^5\slashed{A})\frac{1}{\slashed{p}-m}\slashed{V}\,,
        \\
        \label{Exp4thThetaThetaVA3}
        \ri\tr\frac{1}{\slashed{p}-m}(\ri m\gamma^5\theta)\frac{1}{\slashed{p}-m}\slashed{V}\frac{1}{\slashed{p}-m}(\ri m\gamma^5\theta)\frac{1}{\slashed{p}-m}(-\gamma^5\slashed{A})\,,
    \end{align}
\end{subequations}
in expansion \eqref{GammaExp}, but also the third-order terms
\begin{subequations}
    \begin{align}
        \label{Exp3rdThetaThetaVA1}
        \ri\tr\frac{1}{\slashed{p}-m}\left(-\frac{m}{2} \theta^2 \right)\frac{1}{\slashed{p}-m}\slashed{V}\frac{1}{\slashed{p}-m}(-\gamma^5\slashed{A})\,,
        \\
        \label{Exp3rdThetaThetaVA2}
       \ri\tr\frac{1}{\slashed{p}-m}\left(-\frac{m}{2} \theta^2 \right)\frac{1}{\slashed{p}-m}(-\gamma^5\slashed{A})\frac{1}{\slashed{p}-m}\slashed{V}\,.
    \end{align}
\end{subequations}
After simplification, the resulting effective action is
\begin{equation}
    \label{GammaThetaThetaVA}
    \begin{split}
        \Gamma_{\theta\theta VA}
       =\frac{\ri}{48\pi^2}\epsilon^{\mu\nu\alpha\beta}\int \rd^4x\Tr\Big( 
        2[\partial_{\mu}\theta,\theta]\{\partial_{\nu}V_{\alpha},A_{\beta}\} 
        +2(\partial_{\mu}\theta\partial_{\nu}V_{\alpha}\theta 
        -\theta\partial_{\nu}V_{\alpha}\partial_{\mu}\theta )A_{\beta}
           \\     +\theta^2[\partial_{\nu}V_{\alpha},\partial_{\mu}A_{\beta}]
        \Big)\,,
    \end{split}
\end{equation}
up to first order in derivatives. 
Because this is one of the most involved calculations and the result plays a key role in the applications discussed in the next section, we provide some computational details in Appendix~\ref{app:deriv}. 

Finally, the $\theta VAA$ contribution is also nonzero. It is obtained from terms \eqref{Exp4thThetaThetaVA} with one $\ri m\gamma^5\theta$ replaced by a $-\gamma^5\slashed{A}$ in each line such that each term is a distinct ordering of the external fields under the cyclicity of the trace. This piece of the effective action reads
\begin{equation}
    \label{GammaThetaVAA}
    \begin{split}
        \Gamma_{\theta VAA} = 
        \frac{\ri}{48\pi^2}\epsilon^{\mu\nu\alpha\beta}\int \rd^4x \Tr\Big(
        2\theta\partial_{\mu}V_{\nu}A_{\alpha}A_{\beta}-2\theta V_{\nu}A_{\alpha}\partial_{\mu}A_{\beta}
        +2\partial_{\mu}\theta A_{\nu}V_{\alpha}A_{\beta} 
        \\
        -6\theta A_{\nu}\partial_{\mu}V_{\alpha}A_{\beta}
        -2\theta\partial_{\mu}A_{\nu} A_{\alpha}V_{\beta} +2\theta A_{\nu}A_{\alpha}\partial_{\mu}V_{\beta}
       \Big)
        \,,
    \end{split}
\end{equation}
again up to first order in derivatives.

Collecting all terms, the WZW effective action up to fourth order in $V_{\mu}$, $A_{\mu}$, and $\theta$ is
\begin{equation}
    \label{GammaWZW}
    \begin{split}
        \Gamma_{\text{WZW}}(V,A,\theta) 
        =& 
        \frac{\ri N_c}{48\pi^2}\epsilon^{\mu\nu\alpha\beta}\int \rd^4x \Tr_f \Big(
        -2\ri\partial_{\mu}\theta\left(
        3\partial_{\nu}V_{\alpha}V_{\beta}
        +\partial_{\nu}A_{\alpha}A_{\beta}
        \right)
        \\
        &
        -\partial_{\mu}\theta \partial_{\nu}\theta \partial_{\alpha}\theta V_{\beta}
        +6(
        \partial_{\mu}\theta V_{\nu}V_{\alpha}V_{\beta} 
        +\theta V_{\nu}\partial_{\mu}V_{\alpha}V_{\beta}
        )
        \\
        &+2\theta\partial_{\mu}V_{\nu}A_{\alpha}A_{\beta}
        -2\theta V_{\nu}A_{\alpha}\partial_{\mu}A_{\beta}
         +2\partial_{\mu}\theta A_{\nu}V_{\alpha}A_{\beta}
        \\
        &
        -6\theta A_{\nu}\partial_{\mu}V_{\alpha}A_{\beta}
        -2\theta\partial_{\mu}A_{\nu} A_{\alpha}V_{\beta} 
        +2\theta A_{\nu}A_{\alpha}\partial_{\mu}V_{\beta}
        \\
        &
        +2[\partial_{\mu}\theta,\theta]\{\partial_{\nu}V_{\alpha},A_{\beta}\} 
        +2(\partial_{\mu}\theta\partial_{\nu}V_{\alpha}\theta
        -\theta\partial_{\nu}V_{\alpha}\partial_{\mu}\theta )A_{\beta}
        \\ &+\theta^2[\partial_{\nu}V_{\alpha},\partial_{\mu}A_{\beta}]
        \Big)
        +\dots\,.
    \end{split}
\end{equation}
In the above, we have performed the color trace, leaving only the flavor trace $\Tr_f$. Our result completely agrees with the full non-Abelian anomaly commonly used in ChPT \cite{Bijnens:1994qh,Maiani:1995ve,Kaiser_2000,Scherer:2002tk}. Typically, this effective action is written in terms of differential forms of the left- and right-handed gauge fields and field $U=\exp {(-\ri\theta)}$. It is uncommon to express the WZW term directly in terms of the vector, axial-vector fields and pseudoscalar fields (see Refs.~\cite{Kaiser_2000, Kaiser:2000ck, Fukushima:2012fg} for other examples). The general expression above thus serves as a useful compact form of the full effective action, particularly in contexts where the vector and axial-vector variables are more convenient.

For our purposes, we specialize to the case of an Abelian vector or axial-vector gauge field obeying the transformation \eqref{locax} and decompose $\theta$ according to Eq.~\eqref{decomp}. To this end, let us change notation to $V_{\mu}\rightarrow eQV^Q_{\mu}$ with charge matrix $Q$ in analogy to electromagnetism. The effective action \eqref{GammaWZW} then reduces to
\begin{equation}
    \label{GammaAbel}
    \begin{split}
        \Gamma = \frac{N_c}{8\pi^2}\epsilon^{\mu\nu\alpha\beta}\int \rd^4x\Tr_f\left\{\partial_{\mu}(\Pi+\phi)(e^2Q^2\partial_{\nu}V^Q_{\alpha}V^Q_{\beta} +\frac{1}{3}\partial_{\nu}A_{\alpha}A_{\beta})
         \right.
        \\
        \left.
        -\frac{\ri e}{3}Q\partial_{\mu}\left(\Pi+\phi\right) \partial_{\nu}\Pi \partial_{\alpha}\Pi V^Q_{\beta} 
        +\frac{\ri e}{3}Q [\partial_{\mu}\Pi,\Pi]\partial_{\nu}V^Q_{\alpha}A_{\beta}
        \right\}\,.
    \end{split}
\end{equation}
By looking at the variation of the above under the axial transformation \eqref{locax}, we can find the divergence of the axial current $j_{A}^{\mu}$. To be specific, we use
\begin{equation}
    \label{varGamma}
    \delta_{A}\Gamma 
    = \int \rd^4x \partial_{\mu}\lambda j^{\mu}_{A} 
    = -\int \rd^4x \lambda \partial_{\mu}j^{\mu}_{A}
\end{equation}
where $\delta_{A}$ denotes the variation under the axial transformation \eqref{locax}, to find
\begin{equation}
    \label{divj5}
    \begin{split}
        \partial_{\mu}j^{\mu}_A 
        = \frac{N_c}{4\pi^2}\epsilon^{\mu\nu\alpha\beta}\left(
        \frac{e^2}{4} \Tr_f Q^2 F^Q_{\mu\nu}F^Q_{\alpha\beta}
        +\frac{N_f}{12}F^A_{\mu\nu}F^A_{\alpha\beta} 
        \right)
        \,,
    \end{split}
\end{equation}
where the field strength tensor for $V^Q_{\mu}$ is defined as $F^Q_{\mu\nu}\equiv\partial_{\mu}V^Q_{\nu}-\partial_{\nu}V^Q_{\mu}$ and likewise for the $A_{\mu}$ field strength tensor $F^A_{\mu\nu}$. This matches what one might expect from the Abelian version of the Bardeen anomaly in Ref.~\cite{Bardeen:1969md}.

Up to this point, we have not specified the vector gauge transformation under which our Lagrangian \eqref{lagLSM} is invariant. Using the same substitution $V_{\mu} = eQV^Q_{\mu}$ for an Abelain vector field and the decomposition \eqref{decomp}, it is invariant under the infinitesimal transformation
\begin{equation}
    \label{vec2}
    \begin{split}
        \psi\rightarrow(1-\ri e\chi Q)\psi\,,
        \qquad V^Q_{\mu}\rightarrow V^Q_{\mu}+\partial_{\mu}\chi\,, 
        \qquad \Pi\rightarrow\Pi +\ri e\chi[\Pi,Q]\,.
    \end{split}
\end{equation}
Under this transformation, the variation of the effective action \eqref{GammaAbel} is nonzero. In other words, the equivalent of Eq.~\eqref{varGamma} for transformation \eqref{vec2} is nonzero. However, this variation is fourth order in the fields. We expect the vector current $j^{\mu}_Q$ to be one order lower in the fields than our effective action (i.e., third order). As such, the higher-order contribution to the divergence of $j^{\mu}_Q$ can be ignored for our purposes. Furthermore, one can confirm this variation is cancelled by terms appearing if $\Gamma_{\text{WZW}}$ were extended to one order higher in the external fields. Since such terms are beyond the scope of this work, we do not include them here. It suffices to note that the full effective action is vector gauge invariant up to at least fourth order in the external fields. Thus, we conclude that up to the order of interest, the vector current $j^{\mu}_Q$ is conserved and the consistent anomaly is reproduced.

\section{Pions with vorticity, electromagnetic fields, and chemical potentials}
\label{sec:app}
Here, we shall discuss some of the physical applications of our effective action for $N_f = 2$. Before discussing the vortical effects brought about by $A_{\mu}$, we first focus on the relevance of our results when both $A_{\mu}$ and $\phi$ are absent. Without these fields, we retrieve the usual anomalous coupling between pseudoscalar mesons and electromagnetic fields \cite{Wess:1971yu,Witten:1983tw,Goldstone:1981kk}, including the one which contributes to the neutral pion decay to two photons \cite{Adler:1969gk,Bell:1969ts}. 

In addition, there are anomalous terms which emerge in the presence of background chemical potentials. For this purpose, we also add the term
\begin{equation}
    \label{lagext}
    \begin{split}
        \mathcal{L}_{\text{ext}} = \bar{\psi}_f\gamma^0\mu_f\psi_f\,,
    \end{split}
\end{equation}
to our Lagrangian \eqref{lagLSM}, where $\mu_f$ is the chemical potential associated with flavor $f$. 
By promoting the chemical potentials to $\U(1)$ vector fields, we can essentially make the substitution $V_{\mu}=eQV^Q_{\mu}+V^{B}_{\mu}/N_c+\tau_3V^I_{\mu}/2$ in the effective action \eqref{GammaWZW}, where $V^B_{\mu}$ and $V^I_{\mu}$ are the baryon and isospin gauge fields, respectively, $Q = \rm{diag}(2/3, -1/3)$ is the charge matrix, and $\tau_a$ are Pauli matrices. After carrying out the flavor and color traces, we obtain the following terms in the effective action for $N_c=3$:
\begin{subequations}
    \begin{align}
         \label{GammaPiyB}
        \Gamma_{\pi Q B} + \Gamma_{\pi\pi\pi B} =& \frac{e}{8\pi^2f_{\pi}}\epsilon^{\mu\nu\alpha\beta}\int \rd^4x\,\left(\partial_{\mu}\pi_0 F_{\nu\alpha}^Q
        +\frac{2}{3f_{\pi}^2}\epsilon_{abc}\partial_{\mu}\pi_a\partial_{\nu}\pi_b\partial_{\alpha}\pi_c\right)V^B_{\beta} \,,
        \\
        \label{GammaPiyI}
         \Gamma_{\pi Q I} =& \frac{e}{16\pi^2f_{\pi}}\epsilon^{\mu\nu\alpha\beta}\int \rd^4x\,\partial_{\mu}\pi_0 F_{\nu\alpha}^QV^I_{\beta}\,,
    \end{align}
\end{subequations}
where one can eventually set $V_{\mu}^B=(\mu_B,\bm{0})$ and $V^I_{\mu}=(\mu_I,\bm{0})$ as in Ref.~\cite{Son:2004tq}. The effective Lagrangian extracted from Eq.~\eqref{GammaPiyB} is essentially $V^B_{\mu}j^{\mu}_{\text{GW}}$, where $j^{\mu}_{\text{GW}}$ is a baryon Goldstone-Wilczek current \cite{Goldstone:1981kk,Witten:1983tw}. This is the same lowest-order term in $\pi_a$ derived in Refs.~\cite{Son:2004tq,Son:2007ny}, which was used to obtain the pion domain wall \cite{Son:2007ny}, and was later generalized to the CSL phase~\cite{Brauner:2016pko} in QCD at finite $\mu_B$ in a magnetic field. 
The effective Lagrangian extracted from Eq.~\eqref{GammaPiyI} yields the analogous term for the isospin CSL~\cite{Brauner:2019aid,Gronli:2022cri}. Our results demonstrate that all these terms can be derived explicitly from the linear sigma model.

We now include a constant vorticity $\omega^{\mu}= \epsilon^{\mu \nu \alpha \beta}u_{\nu}\partial_{\alpha} u_{\beta}/2$ with $u^{\mu}$ being four-velocity.
Applying the identification of the vorticity as an emergent axial gauge field $A_{\mu}=\omega_{\mu}/2$ \cite{Yamamoto:2021gts}, together with the observation that the axial current $j^{5\mu} = \bar{\psi}\gamma^5\gamma^{\mu}\psi$ can be interpreted as the spin polarization, the term involving $A_{\mu}$ corresponds to the spin-vorticity coupling $j^{5\mu} \omega_{\mu}/2$ in Lorentz covariant form.
In such a case, the last term in Eq.~(\ref{GammaAbel}) for $\SU(2)$ flavor symmetry leads to an effective action,
\begin{equation}
    \label{pipiQw}
        \begin{split}
            \Gamma_{\pi\pi Q\omega}=-\frac{e}{16\pi^2f_{\pi}^2}\epsilon^{\mu\nu\alpha\beta} \int \rd^4x
            \left(\pi_2\partial_{\mu}\pi_1 -\pi_1\partial_{\mu}\pi_2\right)F^Q_{\nu\alpha}\omega_{\beta}
        \end{split}
\end{equation}
which couples the electromagnetic gauge field and pion fields to a background vorticity. 
This term accounts for an anomalous current and angular momentum density:
\begin{subequations}
\begin{align}
\label{j}
    j^{\mu} &= \frac{1}{e}\frac{\delta \Gamma_{\pi\pi Q\omega}}{\delta V^{Q}_{\mu}}
    = -\frac{\ri}{4\pi^2f_{\pi}^2}\epsilon^{\mu\nu\alpha\beta} \omega_{\nu}\partial_{\alpha}\pi^{+}\partial_{\beta}\pi^{-}\,, \\
    \label{J}
    J^{\mu}&= \frac{\delta \Gamma_{\pi\pi Q\omega}}{\delta \omega_{\mu}}= -\frac{e}{16\pi^2f_{\pi}^2}\epsilon^{\mu\nu\alpha\beta} F_{\nu \alpha}^Q j_{\beta}^I\,,
\end{align}
\end{subequations}
where 
\begin{equation}
    \label{jI}
    j^{\mu}_I = \epsilon^{3 ab} \pi^a \partial^{\mu} \pi^b = -\ri (\pi^+ \partial^{\mu} \pi^{-} - \pi^- \partial^{\mu} \pi^{+})
\end{equation}
is the non-anomalous isospin current carried by charged pions with $\pi^{\pm} \equiv (\pi_1 \mp \ri \pi_2)/\sqrt{2}$ from the lowest-order Lagrangian in ChPT (see, e.g., Ref.~\cite{Scherer:2002tk}).
This is one of our main results. 
In particular, the temporal component of Eq.~(\ref{j}) and spatial component of Eq.~(\ref{J}) are given by
\begin{subequations}
\begin{align}
n &= \frac{\ri}{4\pi^2f_{\pi}^2} {\bm \omega} \cdot ({\bm \nabla} \pi^+ \times {\bm \nabla} \pi^-)\,, 
\\
{\bm J} & = -\frac{e n_I}{8\pi^2f_{\pi}^2} {\bm B}\,,
\end{align}
\end{subequations}
respectively, where $n_I=j^0_I$ is the non-anomalous isospin charge. The former is the anomalous charge carried by pions under the presence of vorticity. The latter is the anomalous angular momentum carried by pions in the presence of the magnetic field ${\bm B}$, which can be viewed as a cross-correlated response between vorticity and magnetic field; see also Ref.~\cite{Huang:2017pqe} for a similar cross-correlation.

Combining with the lowest-order Lagrangian for the electromagnetic interaction term in ChPT, we have the vorticity-modified photon-pion coupling
    \begin{eqnarray}
	   \mathcal{L}^{\gamma\pi\pi}_{(2,\omega)}
       =\ri eV^{Q\mu}\Big(\pi^{+}\partial_{\mu}\pi^{-}-\pi^{-}\partial_{\mu}\pi^{+} -\frac{1}{4\pi^2f_{\pi}^2}\epsilon_{\mu\nu\alpha\beta}\omega^{\nu}\partial^{\alpha}\pi^{+}\partial^{\beta}\pi^{-}\Big)\,.
    \end{eqnarray}
Such an effective interaction is expected to modify the current-current correlators associated with the photon and dilepton production in a pion gas at finite temperature and/or density. 

Furthermore, we can consider the  additional effective action with constant vorticity involving $V^I_{\mu}$:
    \begin{equation}
    \label{pipiIw}
         \Gamma_{\pi\pi I\omega} 
         = -\frac{\ri}{4\pi^2f_{\pi}^2}\epsilon^{\mu\nu\alpha\beta}\int \rd^4 x
         \,V_{\mu}^I\omega_{\nu}\partial_{\alpha}\pi^{+}\partial_{\beta}\pi^{-} 
         \,.
\end{equation}
Both Eqs.~\eqref{pipiQw} and \eqref{pipiIw} enter at fourth order within the momentum power counting scheme $V^Q_{\mu}, V^I_{\mu}, \partial_{\mu}=\mathcal{O}(p^1)$ in ChPT, and are thus subleading. Even so, they are unique in that they couple electromagnetic fields and/or $\mu_I$ with vorticity to charged pions. Therefore, they may be relevant for discussions of charged pion condensation in systems with large rotation, magnetic fields, and/or $\mu_I$, such as heavy ion collisions. In particular, there has been interest in the condensation of charged pions under rotation in a magnetic field \cite{Liu:2017spl,Liu:2017zhl,Chen:2019tcp,Guo:2024ocw}. The $\Gamma_{\pi\pi Q\omega}$ term would affect the dispersion relation of the charged pions, leading to a correction to the effective chemical–potential threshold for condensation. 

\section{Summary and Outlook}
\label{sec:summary}
Starting from a linear sigma model, we have explored some applications of WZW terms involving vector, axial-vector, and pseudoscalar fields, giving particular focus to the implications of vorticity as an emergent axial gauge field proposed in Ref.~\cite{Yamamoto:2021gts}. By first deriving the WZW effective action up to fourth order in fields via a derivative expansion method introduced in Ref.~\cite{Aitchison:1985pp} and confirming the results in ChPT \cite{Bijnens:1994qh,Maiani:1995ve,Kaiser_2000,Scherer:2002tk}, we showed that our effective action reproduces the consistent anomaly for Abelian vector and axial-vector fields. 
We also derived the anomalous terms which couple external baryon and isospin chemical potentials to pions in a magnetic field, which are responsible for the CSL phase in QCD \cite{Son:2007ny,Brauner:2016pko,Gronli:2022cri}.
Through interpreting the axial-vector field as the vorticity, we found electromagnetic and vortical couplings to charged pions. The new terms encompass the anomalous current under the presence of vorticity, the anomalous angular momentum in a magnetic field, and the vorticity-modified photon-pion coupling. The last is expected to affect the current-current correlators relevant to photon and dilepton production in a pion gas, as well as charged pion condensation in the presence of rotation and magnetic fields \cite{Liu:2017spl,Liu:2017zhl,Chen:2019tcp,Guo:2024ocw}.

The primary applications of our results lie in the context of heavy ion collisions. In future work, it would be interesting, e.g., to evaluate the vorticity-induced corrections to the dilepton production rate from a thermal pion gas, complementing those from quark-gluon plasmas in Refs.~\cite{Singh:2018bih,Wei:2021dib,Dong:2021fxn}. 
Since these phenomena occur at nonzero temperatures, our results should be extended to include thermal effects. 

While the HME in Ref.~\cite{Yamamoto:2021gts} can be derived from our WZW terms by following the procedure in Ref.~\cite{Bardeen:1984pm} (see also Ref.~\cite{Landsteiner:2016led}), it is expected from anomaly matching that the HME should emerge also from the low-energy gapless modes once Dirac fermions dynamically acquire a mass. Such terms, however, do not appear in our present analysis. Similarly, we have not obtained the couplings between neutral pions
or $\eta'$ to vorticity at finite $\mu_B$ and/or $\mu_I$, which were obtained via anomaly matching in Refs.~\cite{Huang:2017pqe,Nishimura:2020odq}. 
It would be interesting to establish these effects through an alternative derivation. We leave these questions for future work.

\section*{Acknowledgements}
GE and DL would like to thank Kazuya Mameda for useful discussion. GE would also like to give thanks to Yair Mulian  for useful discussion,  David Wagner  for introducing them to the FeynCalc package as well as the authors of FeynCalc  \cite{Shtabovenko:2023idz,Shtabovenko:2020gxv,SHTABOVENKO2016432,MERTIG1991345}, and Tom\'{a}\v{s} Brauner and the Department of Physics and Mathematics of the University of Stavanger for their hospitality and valuable insights and discussion. This work was supported by National Science and Technology Council (Taiwan) under Grant No. NSTC 113-2628-M-001-009-MY4, Academia Sinica under Project No. AS-CDA-114-M01, JSPS KAKENHI Grant No. JP24K00631, and the Ishii-Ishibashi Fund (Keio University Grant for Early Career Researchers).

\appendix
\section{Details of the derivation of Eq.~(\ref{GammaThetaThetaVA})}
\label{app:deriv}
This appendix includes details of our derivation of $\Gamma_{\theta\theta VA}$ to demonstrate the application of the derivative expansion method of fermion determinants outlined in Sec.~\ref{sec:setup} and developed in Ref.~\cite{Aitchison:1985pp}. We begin from the third-order term in the expansion \eqref{Exp3rdThetaThetaVA1}, which we can write as
\begin{equation}
    \begin{split}
        \label{AppExp3rdThetaThetaVA1}
        \ri\tr\frac{\slashed{p}+m}{X}\left(-\frac{m}{2}\theta^2 \right)\frac{\slashed{p}+m}{X}\slashed{V}\frac{\slashed{p}+m}{X}(-\gamma^5\slashed{A})\,,
    \end{split}
\end{equation}
where $X=p^2-m^2$. Note also that the pseudoscalar fields appear from the second-order term in the expansion $\re^{\ri\theta\gamma^5}-1=\ri\theta\gamma^5 -\theta^2/2!+\dots$ for small $\theta$. By expanding the numerators, we find that the terms proportional to $m^2$ are the only ones with a nonzero Dirac trace. 
When using dimensional regularization, one should evaluate the Dirac trace after performing the momentum integration.
This is due to the ambiguity of defining $\gamma^5$ in $d$ dimensions. We use the interpretation in Ref.~\cite{tHooft:1972tcz} in such situations where divergent momentum integrals require the Dirac trace to be evaluated in $d$ dimensions to find finite results. For ease of reference, here are some general integral results computed via dimensional regularization:
\begin{equation}
    \label{DimReg}
    \begin{split}
        \int \frac{\rd^d p}{(2\pi)^d} \frac{1}{(p^2-m^2+\ri\epsilon)^n} 
        = 
        &\frac{(-1)^n \ri}{(4\pi)^{d/2}}\frac{\Gamma (n-d/2)}{\Gamma(n)}\frac{1}{(m^2)^{n-d/2}}\,,
        \\
         \int \frac{\rd^d p}{(2\pi)^d} \frac{p_{\mu} p_{\nu}}{(p^2-m^2+\ri\epsilon)^n} 
         = 
         &\frac{(-1)^{n-1} \ri}{(4\pi)^{d/2}}\frac{\Gamma (n-1-d/2)}{\Gamma(n)}\frac{1}{(m^2)^{n-d/2-1}}\frac{1}{2}\eta_{\mu\nu}\,,
          \\
         \int \frac{\rd^d p}{(2\pi)^d} \frac{p_{\mu} p_{\nu} p_{\alpha} p_{\beta}}{(p^2-m^2+\ri\epsilon)^n} 
         = &\frac{(-1)^{n-2} \ri}{(4\pi)^{d/2}}\frac{\Gamma (n-2-d/2)}{\Gamma(n)}\frac{1}{(m^2)^{n-d/2-2}}\frac{1}{4}\left( \eta_{\mu\nu}\eta_{\alpha\beta} 
         \right.
         \\
         &\left.
         +\eta_{\mu\alpha}\eta_{\nu\beta} 
         +\eta_{\mu\beta}\eta_{\nu\alpha}\right)\,,
    \end{split}
\end{equation}
where $\Gamma(x)$ is the Gamma function. 
(The integrals with odd number of momenta in the numerator vanish.) One can usually judge whether divergent integrals will be encountered in the calculation simply from power counting in momentum. For the term \eqref{AppExp3rdThetaThetaVA1}, we expect that the momentum integrals in $d=4$ should be finite, so we evaluate the Dirac trace immediately. We have
\begin{equation}
    \begin{split}
    -2m^2\epsilon^{\mu\nu\alpha\beta}\tr\frac{1}{X}\left(p_{\mu}\theta^2\frac{p_{\nu}}{X}V_{\alpha}
    +p_{\mu}\theta^2\frac{1}{X}V_{\nu}p_{\alpha}
    +\theta^2\frac{p_{\mu}}{X}V_{\nu}p_{\alpha}
    \right)\frac{1}{X}A_{\beta}
    \\
    =
    -2m^2\epsilon^{\mu\nu\alpha\beta}\tr\frac{1}{X}\left(2\ri p_{\mu}\theta^2\frac{1}{X}\partial_{\nu}V_{\alpha}
    +\partial_{\mu}(\theta^2)\frac{1}{X}\partial_{\nu}V_{\alpha}
    \right)\frac{1}{X}A_{\beta}
    \,,
    \end{split}
\end{equation}
where we moved all the momentum operators in the numerators to the left in the second line using Eq.~\eqref{Commute1}. Now we also move all factors of $X^{-1}$ to the left. This can be done freely for any terms already of the required order in derivatives. Here, we are only interested in terms up to second order in derivatives. Thus, the only change in the above is
\begin{equation}
    \begin{split}
        \frac{1}{X}p_{\mu}\theta^2\frac{1}{X}\partial_{\nu}V_{\alpha}
        \frac{1}{X}A_{\beta} 
        =
        p_{\mu}\frac{1}{X^3}\theta^2\partial_{\nu}V_{\alpha}
        A_{\beta}
        +
        2\ri p_{\mu}p^{\lambda}\frac{1}{X^4}\left[
        \partial_{\lambda}\left(
        \theta^2\partial_{\nu}V_{\alpha}\right)
        \right.
        \\
        \left.
        +
        \partial_{\lambda}\left(
        \theta^2\right)\partial_{\nu}V_{\alpha}
        \right]A_{\beta} 
        + \mathcal{O}(\partial^3)\,,
    \end{split}
\end{equation}
using the commutation relation \eqref{Commute2}. With our expression in the form of Eq.~\eqref{TrInt}, we can now perform the momentum integrals using Eqs.~\eqref{DimReg} to find
\begin{equation}
    \begin{split}
        \frac{-\ri}{48\pi^2}\epsilon^{\mu\nu\alpha\beta}\int \rd^4x \Tr\,\partial_{\mu}(\theta^2)\partial_{\nu}V_{\alpha}A_{\beta}\,.
    \end{split}
\end{equation}
An almost identical calculation starting from the term \eqref{Exp3rdThetaThetaVA2} yields
\begin{equation}
    \begin{split}
        \frac{\ri}{48\pi^2}\epsilon^{\mu\nu\alpha\beta}\int \rd^4x \Tr\,\partial_{\mu}(\theta^2)\partial_{\nu}A_{\alpha}V_{\beta}\,.
    \end{split}
\end{equation}

The same procedure can be done for the terms \eqref{Exp4thThetaThetaVA}. Let us take term \eqref{Exp4thThetaThetaVA1}, which, similar to before, we write as
\begin{equation}
    \begin{split}
        \ri m^2\tr\frac{\slashed{p}+m}{X}\theta\gamma^5\frac{\slashed{p}+m}{X}\theta\gamma^5\frac{\slashed{p}+m}{X}\slashed{V}\frac{\slashed{p}+m}{X}\gamma^5\slashed{A}\,.
    \end{split}
\end{equation}
There are terms with nonzero Dirac trace proportional to $m^2$ and $ m^4$. Starting with ${\cal O}(m^4)$, we can directly evaluate the Dirac trace and find
\begin{equation}
    \begin{split}
    &4m^4\epsilon^{\mu\nu\alpha\beta}\tr\frac{1}{X}\bigg(
    p_{\mu}\theta\frac{p_{\nu}}{X}\theta\frac{1}{X}V_{\alpha}
    -p_{\mu}\theta\frac{1}{X}\theta\frac{p_{\nu}}{X}V_{\alpha}
    -p_{\mu}\theta\frac{1}{X}\theta\frac{1}{X}V_{\nu}p_{\alpha}
     +\theta\frac{p_{\mu}}{X}\theta\frac{p_{\nu}}{X}V_{\alpha}
    \\
    &
    \quad +\theta\frac{p_{\mu}}{X}\theta\frac{1}{X}V_{\nu}p_{\alpha}
    -\theta\frac{1}{X}\theta\frac{p_{\mu}}{X}V_{\nu}p_{\alpha}
    \bigg)\frac{1}{X}A_{\beta}
    \\
    &=
    -4m^4\epsilon^{\mu\nu\alpha\beta}\tr\bigg(
    \frac{\ri}{X^4}p_{\mu}\left(\partial_{\nu}\theta\theta V_{\alpha}A_{\beta}
    +\theta^2\partial_{\nu}V_{\alpha}
    \right)
    -\frac{2}{X^5}p_{\mu}p^{\lambda}\left[
    \partial_{\lambda}\left(
    \partial_{\nu}\theta \theta V_{\alpha}
    +\theta^2\partial_{\nu}V_{\alpha}
    \right)
    \right.
    \\
    &\left.
    \quad +\partial_{\lambda}\left(
    \partial_{\nu}\theta \theta
    \right) V_{\alpha}    +\partial_{\lambda}\left(\theta^2\right)\partial_{\nu}V_{\alpha}
    +\partial_{\lambda}\partial_{\nu}\theta \theta V_{\alpha}
+\partial_{\lambda}\theta\theta\partial_{\nu}V_{\alpha}
    \right]
    +\frac{1}{X^4}\theta\partial_{\mu}\theta\partial_{\nu}V_{\alpha}
    \bigg)A_{\beta} 
    \\
    &
    \quad +\mathcal{O}(\partial^3)
    \,,
    \end{split}
\end{equation}
where the right-hand side is the result after moving all momentum operators to the left using both commutation relations. After integrating and simplifying, the result at ${\cal O}(m^4)$ is
\begin{equation}
    \begin{split}
        \frac{\ri}{48\pi^2}\epsilon^{\mu\nu\alpha\beta}\int \rd^4x \Tr\left(
        \partial_{\mu}\theta\theta\partial_{\nu}V_{\alpha}
        -\theta\partial_{\mu}\theta\partial_{\nu}V_{\alpha}
        -\partial_{\mu}\theta \partial_{\nu} \theta
        V_{\alpha} 
        \right)A_{\beta}
        \,,
    \end{split}
\end{equation}
for term \eqref{Exp4thThetaThetaVA1}. The other terms \eqref{Exp4thThetaThetaVA2} and \eqref{Exp4thThetaThetaVA3} at ${\cal O}(m^4)$ similarly become
\begin{subequations}
    \begin{align}
        &\frac{\ri}{48\pi^2}\epsilon^{\mu\nu\alpha\beta}\int \rd^4x \Tr\left(\partial_{\mu}\theta\theta\partial_{\nu}A_{\alpha}
        -\theta\partial_{\mu}\theta\partial_{\nu}A_{\alpha}
        +3\partial_{\mu}\theta \partial_{\nu} \theta
        A_{\alpha} 
        \right)V_{\beta}
        \,,
        \\
        &\frac{\ri}{48\pi^2}\epsilon^{\mu\nu\alpha\beta}\int \rd^4x \Tr\left(
        -\partial_{\mu}\theta V_{\alpha}\partial_{\nu}\theta +\theta\partial_{\mu}V_{\alpha}\partial_{\nu}\theta
        +\partial_{\mu}\theta \partial_{\nu}V_{\alpha}  \theta
        \right)A_{\beta}
        \,,
    \end{align}
\end{subequations}
respectively.

At ${\cal O}(m^2)$, we see that we may encounter divergent integrals. Out of this consideration and convenience, we define
\begin{equation}
    \begin{split}
        \mathcal{E}^{\mu\nu\alpha\beta\rho\sigma} \equiv \Tr_{D}\gamma^{\mu}\gamma^5\gamma^{\nu}\gamma^5\gamma^{\alpha}\gamma^{\beta}\gamma^{\rho}\gamma^5\gamma^{\sigma} 
    \end{split}
\end{equation}
and leave its evaluation until later. Applying the commutation relations as before,
\begin{equation}
    \begin{split}
        &\ri m^2\mathcal{E}^{\mu\nu\alpha\beta\rho\sigma} \tr
        \frac{p_{\mu}}{X}\theta\frac{p_{\nu}}{X}\theta \frac{p_{\alpha}}{X}V_{\beta}\frac{p_{\rho}}{X}A_{\sigma} 
        \\
        &=
        m^2\mathcal{E}^{\mu\nu\alpha\beta\rho\sigma} \tr\bigg(
        \frac{1}{X^4}\left[
        p_{\mu}p_{\nu}p_{\alpha}\partial_{\rho}\left(\theta\theta V_{\beta}\right)
        +p_{\mu}p_{\nu}p_{\rho}\partial_{\alpha}\left(\theta\theta\right) V_{\beta}
        +p_{\mu}p_{\alpha}p_{\rho}\partial_{\nu}\theta\theta V_{\beta}\right]
        \\
        &\left.
        \quad +\frac{2\ri}{X^5}p_{\mu}p_{\nu}p_{\alpha}p^{\lambda}\partial_{\rho}\left[\partial_{\lambda}\left(\theta\theta V_{\beta}\right) +\partial_{\lambda}\left(\theta\theta\right)V_{\beta} +\partial_{\lambda}\theta\theta V_{\beta}\right]
        \right.
        \\
        &\left.
        \quad +\frac{2\ri}{X^5}p_{\mu}p_{\nu}p_{\rho}p^{\lambda}\left[\partial_{\lambda}\left(
        \partial_{\alpha}\left(\theta\theta\right) V_{\beta}\right) 
        +\partial_{\lambda}\partial_{\alpha}\left(\theta\theta\right)V_{\beta} +\partial_{\alpha}\left(\partial_{\lambda}\theta\theta\right) V_{\beta}\right]
        \right.
        \\
        &\left.
        \quad +\frac{2\ri}{X^5}p_{\mu}p_{\alpha}p_{\rho}p^{\lambda}\left[
            \partial_{\lambda}\left(
                \partial_{\nu}\theta\theta V_{\beta}
            \right) 
            +\partial_{\lambda}\left(
                \partial_{\nu}\theta\theta
            \right)V_{\beta} 
            +\partial_{\lambda}\left(\partial_{\nu}\theta\right)\theta V_{\beta}
        \right]
        \right.
        \\
        &
        \quad -\frac{\ri}{X^4}\left[
        p_{\mu}p_{\rho}\partial_{\alpha}\left(\partial_{\nu}\theta\theta\right)V_{\beta}
        +p_{\mu}p_{\alpha}\partial_{\rho}\left(\partial_{\nu}\theta\theta V_{\beta}\right)
        +p_{\mu}p_{\nu}\partial_{\rho}\left(\partial_{\alpha}\left(\theta\theta\right) V_{\beta}\right)
        \right]
        \bigg)A_{\sigma}
        \\
        & \quad +\mathcal{O}(\partial^3)\,,
    \end{split}
\end{equation}
where we used $\mathcal{E}^{\mu\nu\alpha\beta\rho\sigma}p_{\mu}p_{\nu}p_{\alpha}p_{\rho} =0$ in $d$ dimensions. 
This removes the only potentially divergent term. We then integrate over the momentum. After contracting the spacetime indices and simplifying the remaining expression, we find
\begin{equation}
    \begin{split}
        \frac{\ri}{48\pi^2}\epsilon^{\mu\nu\alpha\beta}\int \rd^4x \Tr
        \left( \partial_{\mu}\theta\theta \partial_{\nu}V_{\alpha} -\theta\partial_{\mu}\theta\partial_{\nu}V_{\alpha}
        +
        \partial_{\mu}\theta\partial_{\nu}\theta V_{\alpha}
        \right)A_{\beta}\,.
    \end{split}
\end{equation}
In a similar manner, we obtain 
\begin{subequations}
    \begin{align}
        &\frac{\ri}{48\pi^2}\epsilon^{\mu\nu\alpha\beta}\int \rd^4x \Tr
        \left( \partial_{\mu}\theta\theta \partial_{\nu}A_{\alpha} - \theta\partial_{\mu}\theta\partial_{\nu}A_{\alpha}
        +\partial_{\mu}\theta\partial_{\nu}\theta A_{\alpha}
        \right)V_{\beta}\,,
        \\
        &\frac{\ri}{48\pi^2}\epsilon^{\mu\nu\alpha\beta}\int \rd^4x \Tr\left(
        \partial_{\mu}\theta V_{\alpha}\partial_{\nu}\theta
        +\theta\partial_{\mu}V_{\alpha}\partial_{\nu}\theta
        +\partial_{\mu}\theta \partial_{\nu}V_{\alpha}  \theta
        \right)A_{\beta}\,,
    \end{align}
\end{subequations}
at ${\cal O}(m^2)$ from terms \eqref{Exp4thThetaThetaVA2} and \eqref{Exp4thThetaThetaVA3} respectively. 

In total,
\begin{equation}
    \begin{split}
        \Gamma_{\theta\theta VA} = 
        \frac{\ri}{48\pi^2}\epsilon^{\mu\nu\alpha\beta}\int \rd^4x \Tr
        \Big(
        -\partial_{\mu}(\theta^2)\partial_{\nu}V_{\alpha}A_{\beta}
        +\partial_{\mu}(\theta^2)\partial_{\nu}A_{\alpha}V_{\beta}
        \\
        +2[\partial_{\mu}\theta,\theta] \partial_{\nu}V_{\alpha} A_{\beta}  
        +
        2[\partial_{\mu}\theta,\theta] \partial_{\nu}A_{\alpha}V_{\beta} 
        +4\partial_{\mu}\theta\partial_{\nu}\theta A_{\alpha}V_{\beta}
        \\        +2\theta\partial_{\mu}V_{\alpha}\partial_{\nu}\theta A_{\beta}
        +2\partial_{\mu}\theta \partial_{\nu}V_{\alpha}  \theta
        A_{\beta}
        \Big)\,,
    \end{split}
\end{equation}
which, after partially integrating and dropping boundary terms, becomes Eq.~\eqref{GammaThetaThetaVA}.

\interlinepenalty=10000
\raggedbottom
\bibliography{references.bib}

\end{document}